\documentclass[twocolumn]{autart}    

\usepackage{cite}
\usepackage{amsmath,amssymb,amsfonts}
\usepackage{mathrsfs}
\usepackage{cases}  
\usepackage{graphicx}
\usepackage{textcomp}
\usepackage{xcolor}
\usepackage{caption}    

\newtheorem{lemma}{Lemma}
\newtheorem{remark}{Remark}
\newtheorem{definition}{Definition}
\newtheorem{theorem}{Theorem}

\newtheorem{proposition}{Proposition}

\newcommand{\pfbox}{\hfill\mbox{$\Box$}}

\begin{document}
	
	\begin{frontmatter}
		
		\title{Optimal Stationary State Estimation
			Over Multiple \\ Markovian Packet Drop Channels} 
		\thanks[footnoteinfo]{The material in this paper was not presented at any conference.}
		
		\author[ECUST Laboratory]{Jiapeng Xu}\ead{jpxu@mail.ecust.edu.cn},    
		\author[LSU]{Guoxiang Gu}\ead{ggu@lsu.edu},
		\author[ND]{Vijay Gupta}\ead{vgupta2@nd.edu},
		\author[ECUST Laboratory]{Yang Tang}\ead{yangtang@ecust.edu.cn}
		\address[ECUST Laboratory]{Key Laboratory of Advanced Control and Optimization for Chemical Processes, Ministry of Education, East China University of Science and Technology, Shanghai 200237, China.}  
		\address[LSU]{ School of Electrical Engineering and
			Computer Science, Louisiana State
			University, Baton Rouge, LA 70803,
			USA.}        
		\address[ND]{ Department of Electrical Engineering, University
			of Notre Dame, Notre Dame, IN 46556, USA.}


		\begin{abstract}
			In this paper, we investigate
			the  state estimation problem over multiple Markovian packet drop
			channels. In this problem setup,
			a remote estimator
			receives measurement data
			transmitted from multiple sensors over
			individual channels. By the
			method of Markovian jump linear
			systems, an optimal stationary
			estimator that minimizes the error variance in the steady state is obtained, based on the mean-square
			(MS) stabilizing solution to the
			coupled algebraic Riccati equations. An explicit necessary and sufficient condition is derived
			for the existence of the MS
			stabilizing solution, 
			which coincides with that of the
			standard Kalman filter. More
			importantly, we provide a sufficient condition
			under which the MS detectability with multiple
			Markovian packet drop channels can be
			decoupled, and propose a locally optimal stationary estimator but computationally more tractable.
			Analytic sufficient
			and necessary MS detectability
			conditions are presented for the decoupled subsystems subsequently. Finally, numerical simulations are conducted to illustrate the results on the MS stabilizing solution, the MS detectability, and the  performance of the optimal and locally optimal stationary estimators.
		\end{abstract}	
		
		
	\end{frontmatter}
	
	\section{Introduction}	
	
	Networked control systems (NCSs)
	attract a great deal of attention
	from the control community due to
	their numerous advantages over
	conventional control systems.
	Much effort has been devoted to the
	study of control and estimation for
	NCSs over various communication
	channels in recent years. In the case
	of wireless channels, one major issue
	is the occurrence of data packet drops
	that may destroy the feedback stability
	of estimators and controllers. Hence,
	a large number of existing works
	are focused on the stability and
	stabilization of dynamic systems
	over packet drop channels \cite{sinopoli2004kalman,gupta2007optimal,gu2008networked,xiao2012feedback,zhou2019distributed,xu2019optimal,xu2020distributed,tang2020input}.
	
	From the stochastic point of view, the packet drop channel is commonly modeled as either an
	independent and identically distributed
	(i.i.d.) random process or a two-state Markov
	chain by taking the temporal correlation
	into consideration. Under such two modeling
	methods, the stability of Kalman filtering with intermittent measurements has been well studied. In \cite{sinopoli2004kalman}, the authors show that over an i.i.d. packet drop channel, there exists a critical packet arrival rate below which the Kalman filter is unstable. Multi-sensor and distributed scenarios are further studied in \cite{yang2018multi} and \cite{zhou2019distributed}, respectively. Compared with the i.i.d. case, the stability problem of Kalman filtering
	with Markovian packet drops is more
	complicated, yet many interesting
	results are obtained.
	Authors of \cite{huang2007stability}
	are the first to introduce the notion of peak
	covariance to evaluate the estimation performance. Some improved results are obtained in \cite{xie2008stability}, in comparison with \cite{huang2007stability}.
	By showing the equivalent stability property
	for the estimation error covariances
	at packet reception times and each time instant,
	the stability for Kalman filtering has been sufficiently studied in \cite{you2011mean} through exploiting the system structure. In \cite{sui2015stability}, a necessary and sufficient condition is provided for diagonalizable systems with multiple sensors. Similar to
	the existence of the critical packet arrival
	rate shown in the i.i.d. case, the existence
	of the critical curve in terms of the
	failure-recovery rate is proved in
	\cite{wu2018kalman}. For the NCS without an acknowledgment signal sent by the actuator to the estimator, authors of \cite{lin2019state} derive the optimal and an approximate optimal estimator, and show the same stability for both of them.
	
	Since NCSs over Markovian packet drop
	channels can be considered
	as a class of Markov jump
	linear systems (MJLSs)\cite{costa2005discrete}, an alternative approach for studying the state estimation with Markovian packet drops is design
	of optimal stationary jump estimators \cite{smith2003estimation,fletcher2004estimation}, instead of {the} time-varying Kalman filter (TVKF) mentioned above. Although
	the TVKF is known to be the  optimal linear estimator, it does not converge in the steady state and its estimation
	gain explicitly depends on the realization of packet drops such that it needs to be
	computed online.
	In this paper, we consider the stationary state estimation problem over multiple Markovian packet drop channels. Different from the TVKF studied in
	the existing literature, e.g. \cite{sinopoli2004kalman,huang2007stability,xie2008stability,you2011mean,wu2018kalman}, we are interested in the optimal stationary linear state estimator that
	remained unknown for which
	the estimator gains can be
	computed off-line, leading to a reduction in computational burden for the estimator. Compared to \cite{smith2003estimation}, where
	a jump estimator has been designed
	based on the last finite measurement
	loss modes for a single Markovian packet
	drop channel, we consider a nontrivial case of multiple Markovian packet drop channels. More importantly, we investigate two fundamental stability issues, the existence of the MS stabilizing solution to the corresponding coupled algebraic Riccati equations ({CAREs}) and the MS detectability
	for the NCS with Markovian packet drops, which have
	not been studied by \cite{smith2003estimation}.
	
	It is well known that estimation and control
	are dual problems. Their optimal solutions
	are associated with their respective
	Riccati equations. Therefore, 
	feedback control problems over packet drop
	or fading channels are also related to
	the study in this paper. In \cite{schenato2007foundations}, it is shown
	that for the optimal linear quadratic
	Gaussian (LQG) control with i.i.d.
	packet drops, the separation principle
	still holds under a TCP-like protocol,
	and there exists a critical arrival
	probability for the control data. The case
	of multiple lossy channels is further
	considered in \cite{garone2012lqg}, while
	the LQG control with Markovian packet
	drops is studied in \cite{mo2013lqg}.
	In the presence of i.i.d.
	fading channels, the existence of the MS
	stabilizing solution to the modified
	algebraic Riccati equation is studied
	recently in \cite{zheng2018existence}.
	Compared to \cite{zheng2018existence},
	the existence of the MS stabilizing
	solution in this paper is more involved,
	due to the temporal
	correlation of packet drops.
	
	The contributions of this paper are
	summarized as follows.
	\begin{enumerate}
		\item An optimal
		stationary state estimator is obtained
		for the NCS over multiple Markovian
		packet drop channels, by making use of
		the MJLS method. This is a nontrivial generalization from a single channel studied in \cite{smith2003estimation}, since our case brings a challenging issue that the complexity of the optimal estimator increases exponentially with respect to the number of channels.
		\item A necessary and sufficient
		condition is derived for the existence of
		the MS stabilizing solution to the
		associated filtering {CAREs}.
		It is shown
		that, in addition to the MS detectability,
		only the controllability of
		the eigenvalues on the unit circle
		is required, which is weaker than
		the stabilizability assumption in e.g.
		\cite{sinopoli2004kalman,zhou2019distributed,you2011mean,smith2003estimation}.
		\item We provide some sufficient
		and necessary conditions for the MS
		detectability and propose a locally optimal stationary estimator, through exploring the system
		structure. Specifically,
		the MS detectability with multiple
		Markovian packet drop channels can
		be decoupled and a locally optimal estimator, lowering
		the complexity from $N=2^m$ to
		$2m$, is obtained,
		resulting in significant reduction
		on the computational complexity. Moreover,
		some analytic MS detectability conditions
		are derived for the decoupled subsystems.
	\end{enumerate}
	
	The remainder of this paper is organized as follows. Section \ref{sec-estimator} describes the problem considered in this paper and gives the optimal stationary estimator. Section \ref{sec-existence} is focused on the existence of the MS stabilizing solution, which is hinged on the MS detectability and the
	controllability of the eigenvalues on the unit circle. The MS detectability is studied, and a locally optimal stationary estimator is proposed in Section \ref{sec-det}. Numerical examples are provided in Section \ref{sec-examples} to illustrate the stability results and the performance of the proposed estimators, followed by some concluding remarks in Section \ref{sec-con}.

	The notations in this paper are standard. $\mathbb{R}^{m\times n}$ ($\mathbb{C}^{m\times n}$ respectively) denotes the set of $m\times n$ real (complex) matrices, with $\mathbb{R}^{n} := \mathbb{R}^{n\times 1}$ ($\mathbb{C}^{n} := \mathbb{C}^{n\times 1}$). $\mathbb{S}^n_+$ is the set of $n\times n$ real positive semidefinite matrices. For a matrix or vector $X$, denote by $X^*, X'$, and $\bar X$ {the} conjugate  transpose, transpose and conjugate  of $X$, respectively. $\rho(\cdot)$ denotes the spectral radius {of} a matrix or operator, $\rm {diag}\{\cdot\}$ the (block) diagonalization operation, $\rm vec\{\cdot\}$ the vectorization operation, and $\rm {tr}(\cdot)$ the trace of a square matrix. $\otimes$ represents the Kronecker product. $I_n$ represents the identity matrix of dimension $n\times n$ and $\mathbf{1}$ denotes the indicator function. Finally, the expectation operator and the probability of a random event are denoted by $\mathbb{E}[\cdot]$ and $\Pr\{\cdot\}$, respectively.  Other notations will be made clear as we proceed.

	\section{Optimal Stationary State Estimator}\label{sec-estimator}
	Consider {a} discrete-time
	shift-invariant system
	described by
	
	\begin{subequations}\label{xyk}
		\begin{align}
			x(k+1) &= Ax(k) + w(k),\ \ \ \
			x(0)=x_0, \label{xk}\\ 
			y_i(k) &= C_ix(k) + v_i(k),\
			\ \ 1\leq i\leq m, \label{yk}
		\end{align}
	\end{subequations}	
	with $x(k) \in \mathbb{R}^{n}$
	the system state, and $y_i(k)
	\in\mathbb{R}$ the output
	measurement obtained by the
	$i$th sensor.
	$w(k)$ and $v(k)={\rm vec}\{v_1(k),\ldots,
	v_m(k)\}$ are mutually independent white noises
	having mean zero and covariances
	$Q\ge 0$ and $R > 0$, respectively.
	The initial state $x_0$ is independent of $w(k)$
	and $v(k)$, with mean
	$\overline x_0$ and covariance
	$\Pi_0$. Define $y(k):={\rm
		vec}\{y_1(k),\ldots,y_m(k)\}$ and ${C:=[
			C'_1, \ldots, C'_m]'}$.
	We assume without loss of
	generality that $R$ is a diagonal
	matrix.
	
	\begin{remark}
		It is worth mentioning that dimension one
		of the output measurement from each sensor
		does not pose constraints on the results
		in this paper from being generalized to the
		case of arbitrary dimensions. Since our
		focus is on the multiple
		Markovian packet drop processes,
		we consider the collective
		measurement $y(k)\in\mathbb{R}^{m}$
		for convenience. \pfbox
	\end{remark}
	
\begin{figure}[!]
		\centering
		\includegraphics[scale=0.48]{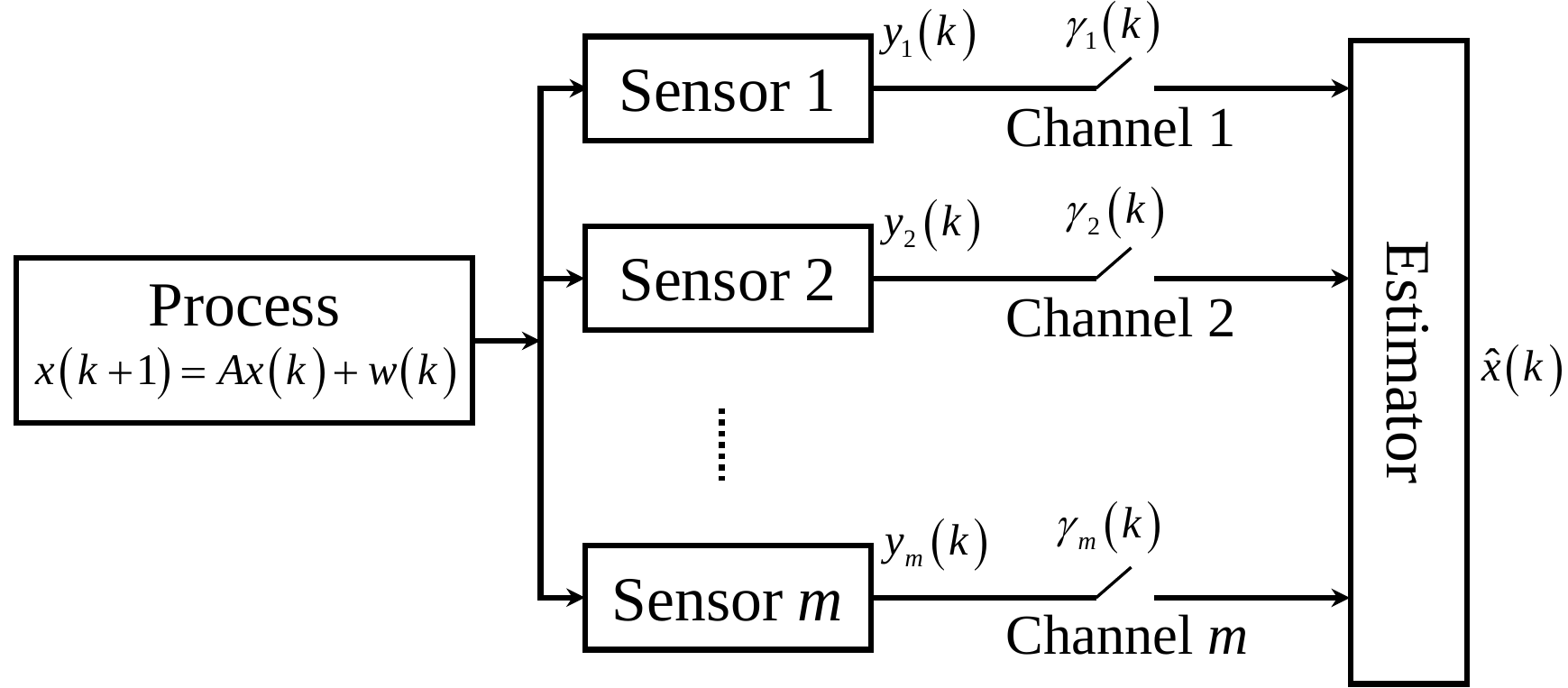}
		\caption{State estimation over packet drop channels.}
		\label{fig_diagram}
	\end{figure}

	We assume that the collective
	measurement $y(k)$ is sent
	through unreliable channels
	suffering from packet drops; see Fig. \ref{fig_diagram}.  The signal received at
	the remote estimator is given by
	\begin{align}\label{gammayk}
		{y_{\rm r}(k)} = \Gamma (k)y(k),
	\end{align}
	where $\Gamma (k)\in
	\mathbb{R}^{m\times m}$
	represents the presence of
	the $m$ packet drop channels
	in the diagonal form:
	\begin{align}
		\Gamma(k) = {\rm diag}\{\gamma_1(k), \gamma_2(k),\ldots,\gamma_m(k)\}.
	\end{align}
	Here, $\gamma_i(k)\in\{0,1\}$
	for $1\leq i\leq m$. If
	$\gamma_i(k) = 1$, $y_i(k)$
	arrives at the estimator;
	otherwise $y_i(k)$ is dropped.
	Moreover, $\{\gamma_i(k)\}_{i=1}^m$
	are independent of each other,
	and each $\gamma_i(k)$ is modeled
	as a time-homogeneous two-state
	Markov chain with the transition
	probability matrix (TPM)
	\begin{align}\label{tpm}
		P_i  = \left[\begin{array}{cc}
			1-q_i & q_i\\ p_i & 1-p_i
		\end{array}\right],\ \
		i=1,\ldots, m,
	\end{align}
	where $q_i = \Pr\{\gamma_i(k+1)=1|\gamma_i(k)=0\}$
	is the recovery rate and $p_i = \Pr\{\gamma_i(k+1)=0|\gamma_i(k)=1\}$
	is the failure rate. Denote
	$\pi_{i,l}(k) := \Pr\{\gamma_{i}(k)={l-1}\}, ~l\in\{1,2\}$. Assume that
	$0<p_i,q_i< 1$. Then, there
	exists the limit probability
	distribution
	$\pi_{i} = \left[\ \pi_{i,1}~
	\pi_{i,2}\ \right]$ with
	\[
	\pi_{i,1} = \frac{p_{i}}{p_{i} + q_{i}},\ \ \
	\pi_{i,2} = \frac{q_{i}}{p_{i} + q_{i}}.
	\]
	
	Different from the TVKF with Markovian packet drops studied in
	the existing literature \cite{huang2007stability,xie2008stability,you2011mean,sui2015stability,wu2018kalman,lin2019state}, we are interested in the optimal stationary linear state estimator for which
	the estimator gains can be
	computed off-line. We will
	apply the filtering theory from
	MJLSs \cite{costa2005discrete}
	to derive such an estimator.
	It is noted that the NCS
	described in (\ref{xyk}) and
	(\ref{gammayk}) can be rewritten
	into the following jump form:
	\begin{subequations}\label{jumpxyk}
		\begin{align}\label{jumpxyk-a}
			x(k+1) &= Ax(k) + w(k),\
			\ \ x(0)=x_0, \\
			{y_{\rm r}}(k) &= H_{\theta(k)}x(k) + D_{\theta(k)}v(k),
		\end{align}
	\end{subequations}
	where $\theta(k) \in \mathcal{N}
	:= \{1, \ldots, 2^m\}$, specified by
	\begin{align}\label{theta}
		\theta(k) = 1 + \sum_{i=1}^{m}
		2^{i-1}\gamma_i(k),
	\end{align}
	is the Markovian jump variable.
	It follows that $H_{\theta(k)} =
	\Gamma(k)C$ and $D_{\theta(k)} = \Gamma(k)$. For simplicity, denote
	$N = 2^m$ in the remainder of the paper.
	
	\begin{lemma}\label{le-tpm}
		{Given the way of computing $\theta(k)$ in (\ref{theta}), the
			TPM $P$ for the Markov process $\{\theta(k)\}$ is
			given by
			\begin{align}\label{tpmP}
				P := P_m\otimes P_{m-1}\otimes\cdots\otimes
				P_{1},\ \ \ 1\leq i,j\leq 2^m,
			\end{align}
			where $P=\left[\ p_{ij}\ \right]$,  $p_{ij}=\Pr\{\theta(k+1)=j|\theta(k)=i\}$.} Moreover, by defining $\mu_i(k) :=
		\Pr\{\theta(k) = i\}$ for each
		$i\in\mathcal{N}$, and $\mu(k) :=
		\left[\ \mu_1(k)\ \cdots\ \mu_N(k)\
		\right]$, $\theta(k)$ has a unique stationary distribution $\mu = \left[\ \mu_1\
		\cdots\ \mu_N\ \right]$, i.e.,
		$\lim_{k\to \infty} \mu_i(k) = \mu_i$ for $i\in\mathcal{N} $ and
		\begin{align}\label{tpmmu}
			\mu = \pi_{m}\otimes\pi_{m-1}\otimes\ldots\otimes\pi_1.
		\end{align}
	\end{lemma}
	
	{\noindent\textbf{Proof.}
		We will prove that (\ref{tpmP}) holds for any $m\ge 1$ by induction. Obviously, (\ref{tpmP}) holds when $m=1$. Assume that for $m=l$ satisfying $l>1$, we have
		\begin{align}
			P^{\{l\}} := \left[\ p_{ij}^{\{l\}}\
			\right] = P_l\otimes\cdots\otimes
			P_{1},\ \ \ 1\leq i,j\leq 2^l.
		\end{align}
		In this case, denote $\vartheta(k)$ as the Markovian jump variable such that $p_{ij}^{\{l\}}=
		\Pr\{\vartheta(k+1)=j|\vartheta(k)=i\}$.
		For $m=l+1$, suppose that sensor $l+1$ is the newly added sensor compared with the case $m=l$. Define
		\begin{align}
			P^{\{l+1\}}: = \left[\
			p_{rs}^{\{l+1\}}\ \right],\ \ \ 1\leq r,s\leq 2^{l+1}
		\end{align}
		as the corresponding TPM. Then by (\ref{theta}) and the assumption that $\{\gamma_i(k)\}_{i=1}^m$ are independent,
		\begin{align}
			&\Pr\{\gamma_{l+1}(k+1),\vartheta(k+1)=j|\gamma_{l+1}(k),\vartheta(k)=i\}\notag\\
			=& p_{(2^l\gamma_{l+1}(k)+i)(2^l\gamma_{l+1}(k+1)+j)}^{\{l+1\}}\notag\\
			=& \Pr\{\gamma_{l+1}(k+1)|\gamma_{l+1}(k)\}\Pr\{\vartheta(k+1)=j|\vartheta(k)=i\},\notag\\
			&~\gamma_{l+1}(k+1),\gamma_{l+1}(k)\in\{0,1\},\ 1\leq i,j\leq 2^l.
		\end{align}
		Therefore, we conclude that
		\begin{align}
			P^{\{l+1\}} = P_{l+1}\otimes P^{\{l\}} = P_{l+1}\otimes P_l\otimes\cdots\otimes
			P_{1}.
		\end{align}
		Similarly, there holds
		\begin{align}
			\mu(k) =
			\pi_{m}(k)\otimes\pi_{m-1}(k)\otimes\ldots\otimes\pi_1(k).
		\end{align}
		Thus, $\mu = \lim_{k\to \infty} \mu(k) = \pi_{m}\otimes\pi_{m-1}\otimes\ldots\otimes\pi_1$. The uniqueness follows from the limit probability distributions $\{\pi_i\}_{i=1}^m$. The proof is thus complete. \hfill\rule{2mm}{2mm}}
	
	Let $\hat x(k)$ be the state
	estimation for $x(k)$ in
	(\ref{jumpxyk-a}). Now, consider
	a dynamic Markovian
	jump linear estimator
	described by
	\begin{align}\label{hatx}
		\hat x(k+1) = A\hat x(k) +
		L_{\theta(k)}[y_{\rm r}(k) -
		H_{\theta(k)}\hat x(k)].
	\end{align}
	At each time instant, the estimator gain $L_{\theta(k)}$ is chosen from a
	finite set of pre-computed values, i.e.,
	$L_{\theta(k)} \in
	\{L_i\in\mathbb{R}^{n\times m}\}_{i\in\mathcal{N} }$.
	This implies that the estimator
	gain $L_{\theta(k)}$ depends only
	on $\theta(k)$ (rather than on
	all the past modes $\{\theta(0),\ldots, \theta(k)\}$, corresponding to the TVKF), which is an important
	feature of the jump estimator.
	We aim to find a set of optimal
	gains, denoted by
	$\{K_i\in\mathbb{R}^{n\times m}\}_{i\in\mathcal{N}}$ such that
	{with
		$L_i=K_i$ for each 
		$i\in\mathcal{N}$}, the 
	stationary estimation cost
	\begin{align}
		J(\infty) = \lim_{k\to
			\infty}\mathrm{E}[\|x(k)
		- \hat x(k)\|^2]
	\end{align}
	is bounded and minimized.
	
	\begin{theorem}\label{th-estimator}
		For the system dynamics described in (\ref{jumpxyk}), assume that there exists the MS stabilizing solution 
		(see Definition \ref{def-stasolution} in Section \ref{sec-existence})
		$Y = (Y_{1},\ldots,Y_{N})$ to the following  {CAREs}
		\begin{align}\label{CARE}
			Y_j = \sum_{i=1}^{N}p_{ij}\big[&AY_{i}A' -  AY_{i}H'_i(H_iY_{i}H'_i + \mu_{i}R)^{-1}H_iY_{i}A' \notag\\
			&+ \mu_{i}Q\big],\ \ \ j\in\mathcal{N}.
		\end{align}
		Then the optimal stationary gains
		for the jump estimator in (\ref{hatx})
		are given by
		\begin{align}\label{Kj}
			K_j = K_j(Y):= AY_{j}H'_j(H_jY_{j}H'_j
			+ \mu_{j}R)^{-1} 
		\end{align}
		for $j\in\mathcal{N} $, and the
		stationary optimal cost is given by
		$$
		\lim_{k\to \infty}J(k)=
		\sum_{i=1}^{N}{\rm tr}(Y_{i}).
		$$
	\end{theorem}
	
	\noindent\textbf{Proof.}
	{The result can be obtained by using the filtering theory of MJLSs (Chapter 5 in  \cite{costa2005discrete}). To be more clear, define the estimation error by $e(k) := x(k) - \hat x(k)$  and set $Y_j(k) := \mathrm{E}\left[e(k)e(k)'\mathbf{1}_{\theta(k)=j}\right], j\in\mathcal{N}$ such that
		\begin{align*}
			\mathrm{E}\left[e(k)e(k)'\right] = \sum_{j=1}^{N}Y_{j}(k).
		\end{align*}
		From (\ref{jumpxyk}) and (\ref{hatx}) we have
		\begin{align}
			e(k+1)=&(A-L_{\theta(k)}(k)H_{\theta(k)})e(k) + w(k) \notag\\
			&- L_{\theta(k)}(k)D_{\theta(k)}v(k)
		\end{align}
		with $L_{\theta(k)}$ replaced by $L_{\theta(k)}(k)$. From Proposition 3.35 2) in \cite{costa2005discrete}, we have
		\begin{align}\label{Yjk}
			Y_j(k+1) =& \sum_{i=1}^{N}p_{ij}[(A - L_{i}(k)H_{i})Y_i(k)(A - L_{i}(k)H_{i})'\notag\\
			 + \mu_{i}&(k)(Q + L_{i}(k)D_{i}RD'_iL'_i(k)],\ j\in\mathcal{N}.
		\end{align}
		By solving $\partial\text{tr}({Y_j(k+1)})/\partial{L_i(k)} = 0$ for all $i\in\mathcal{N}$, it gives $L_i(k) = K_i(k)$ with
		\begin{align}
			K_i(k) = AY_i(k)H_i'(H_iY_i(k)H'_i + \mu_{i}(k)D_iRD'_i)^{\dagger},
		\end{align}
		which minimizes $\text{tr}(Y_j(k+1))$. The Moore-Penrose inverse is used, since matrix $H_iY_{i}H'_i + \mu_{i}(k)D_iRD'_i$ in general is positive semi-definite and the relation ${\mathcal R}\{H_iY_i(k)A'\}\subseteq {\mathcal R}\{H_iY_{i}H'_i + \mu_{i}(k)D_iRD'_i\}$ holds for any $i\in\mathcal{N}$ (${\mathcal R}\{\cdot\}$
		denotes the range space). With $L_i(k)=K_i(k)$, (\ref{Yjk}) becomes
		\begin{align}\label{CDRE}
			Y_j(k+1) = &\sum_{i=1}^{N}p_{ij}\big[AY_{i}(k)A' -  AY_{i}(k)H'_i(H_iY_{i}(k)H'_i \notag\\
			+ &\mu_{i}(k)D_iRD_i')^{\dagger}H_iY_{i}(k)A' + \mu_{i}(k)Q\big].
		\end{align}
		As $k\to\infty$, (\ref{CDRE}) converges to the CAREs (\ref{CARE}), $Y_{j}(k+1)\to Y_{j}$, and $K_i(k)\to K_i$, where $(H_iY_{i}H'_i + \mu_{i}R)^{-1}$ takes place of  $(H_iY_{i}H'_i + \mu_{i}D_iRD'_i)^{\dagger}$ without changing the values of $Y_j$ and $K_i$, following from the special form of $\Gamma(k)$.  Also, the stationary optimal cost is $\sum_{i=1}^{N}\text{tr}(Y_{i})$.}
	\hfill\rule{2mm}{2mm}
	
	\begin{remark}
		It is seen from Theorem \ref{th-estimator} that there are $N=2^m$ CAREs and gains for the optimal stationary estimator, which may cause a difficulty on implementation  when the number of sensors is large. We will deal with this issue in Section \ref{sec-det} by proposing a locally optimal stationary estimator.\pfbox
	\end{remark}
	\section{MS Stabilizing Solution}\label{sec-existence}
	
	From Theorem \ref{th-estimator},
	it is known that the optimal
	stationary estimator is based
	on the existence of the MS
	stabilizing solution to the
	{CAREs} (\ref{CARE}).  Therefore,
	we focus on the necessary and
	sufficient condition for this
	existence problem in this section.
	
	\vspace*{-2mm}
	
	\subsection{Preliminaries of MJLSs}
	
	Let $\mathbb{H}^{m,n}$ represent
	the linear space composed of all
	${N}$-sequences of real matrices
	$V=(V_1,\ldots,V_{N})$ with
	$V_i\in\mathbb{R}^{m,n}$. In the
	case of $m=n$, we denote
	$\mathbb{H}^{n}=\mathbb{H}^{n,n}$,
	and define
	\begin{align*}
		{\mathbb{H}^{n*}}&{:=\{V=(V_1,\ldots,V_{N})\in\mathbb{H}^{n};V_i=V'_i\
			\forall i\in \mathcal{N}\},}\\
		\mathbb{H}_+^{n}&:=\{V=(V_1,\ldots,V_{N})
		\in\mathbb{H}^{n*}; V_i\geq 0 \
		\forall i\in \mathcal{N} \}.
	\end{align*}
	{For
		$V=(V_1,\ldots,V_{N})\in
		\mathbb{H}^{n*}$ and $S=(S_1,\ldots,S_{N})\in
		\mathbb{H}^{n*}$, write that $V\ge S$ (or $V>S$) if $V-S=(V_1-S_1,\ldots,V_N-S_N)\in\mathbb{H}^n_+$ (or $V_i-S_i>0$).}
	It is known that $\mathbb{H}^{m,n}$ can be equipped with the inner product:
	\begin{align}\label{inner}
		\langle V, S\rangle =
		\sum_{i=1}^{N}{\rm tr}(V'_iS_i),
	\end{align}
	where $V=(V_1,\ldots,V_N)$ and
	$S=(S_1,\ldots,S_N)$ are in
	$\mathbb{H}^{m,n}$. For
	$V=(V_1,\ldots,V_{N})\in
	\mathbb{H}^{n}$ and $L=(L_1,\ldots,L_{N})\in
	\mathbb{H}^{n,m}$, define
	the operators $\widetilde{\mathcal{L}}(\cdot) =
	(\widetilde{\mathcal{L}}_1(\cdot),\ldots,\widetilde{\mathcal{L}}_{N}(\cdot))$
	and $\mathcal{L}(\cdot)=
	(\mathcal{L}_1(\cdot),\ldots,\mathcal{L}_{N}(\cdot))$ as
	\begin{align}
		\widetilde{\mathcal{L}}_j(V)
		&:= \sum_{i=1}^{{N}}
		p_{ij}(A+L_iH_i)V_i(A+L_iH_i)',
		\\ \label{tpmlj}
		\mathcal{L}_j(V) &:=
		\sum_{i=1}^{{N}}p_{ij}AV_iA',\ \ \
		j\in\mathcal{N}.
	\end{align}
	Their respective adjoint operators
	$\widetilde{\mathcal{L}}^*$
	and $\mathcal{L}^*$ are given by
	\begin{align}
		\widetilde{\mathcal{L}}_i^*(V) &= \sum_{j=1}^{N}p_{ij}
		(A+L_iH_i)'V_j(A+L_iH_i),
		\\ \label{Listar}
		\mathcal{L}_i^*(V) &= \sum_{j=1}^{N}p_{ij}A'V_jA,\
		\ \ i\in \mathcal{N} ,
	\end{align}
	satisfying the following equalities
	\begin{subequations}\label{adjointinner}
		\begin{align}
			\langle \widetilde{\mathcal{L}}(V),
			S\rangle &= \langle V, \widetilde{\mathcal{L}}^*(S)\rangle, \\
			\langle {\mathcal{L}}(V), S\rangle &=
			\langle V, {\mathcal{L}}^*(S)\rangle.
			\label{adjoint}
		\end{align}
	\end{subequations}
	Recall the inner product defined in
	(\ref{inner}). Denote
	\begin{align*}
		\mathbf A &:= (\underbrace{A,\ldots, A}_N),\ \mathbf Q := (\underbrace{Q,\ldots, Q}_N),\\
		\mathbf H &:= (H_1,\ldots, H_{N}),\ \mathbf p := \{p_{ij}\},\ i,j\in \mathcal{N} .
	\end{align*}
	Next, we introduce several notions regarding MJLSs.
	
	\begin{definition}[Def. 1, \cite{costa1995discrete}]\label{def-det}
		The system described in
		(\ref{jumpxyk}), or simply $(\mathbf H, \mathbf A,\mathbf p)$ is said to be MS detectable, if there exists $L=(L_1,\ldots,L_{N})\in\mathbb{H}^{n,m}$  such that  $\rho(\widetilde{\mathcal{L}})<1$. 
	\end{definition}
	
	\begin{definition}[Def. 5.7, \cite{costa2005discrete}]\label{def-stasolution}
		$Y=(Y_1,\ldots,Y_{N}) \in{\mathbb H}_+^n$ is said to be the MS stabilizing solution for the {CAREs} (\ref{CARE}), if  $\rho(\widetilde{\mathcal{L}})<1$  holds with $L_i = -K_i(Y)$ for
		$i\in\mathcal{N}$.
	\end{definition}
	
	Following the definition of
	the observability for MJLSs
	in Theorem 3 of \cite{shen2013observability},
	we introduce the following
	definition for the uncontrollable eigenvalue of the MJLS in
	(\ref{jumpxyk}). 
	
	\begin{definition}\label{def-uncontr}
		A real number $\lambda\ge 0$
		is said to be  an uncontrollable
		eigenvalue for the pair $(\mathbf A, \mathbf Q)$ if there exists an eigenvector $V=(V_1,\ldots,V_N) \in{\mathbb H}_+^n\backslash\{0\}$ of $\mathcal{L}^*$ such that
		\begin{align}\label{uncontr}
			{\rm (a)} \
			\mathcal{L}^*(V) = \lambda V,\ \
			\ {\rm (b)} \ QV_i = 0\ \ \forall i\in\mathcal{N}.
		\end{align}
	\end{definition}
	
	\vspace*{-2mm}
	
	\subsection{Existence of MS Stabilizing Solution}
	{We begin with the notion of the maximal solution. For $X=(X_1,\ldots,X_{N})\in
		\mathbb{H}^{n*}$, denote $\mathcal{P}_j(X)=\sum_{i=1}^{N}p_{ij}X_j$ and define the operators $\mathcal{X}(\cdot)=(\mathcal{X}_1(\cdot),\ldots,\mathcal{X}_N(\cdot))$ and $\mathcal{R}(\cdot)=(\mathcal{R}_1(\cdot),\ldots,\mathcal{R}_N(\cdot))$ as
		\begin{align}
			\mathcal{X}_j(X)&:=\left[\begin{array}{cc}
				A\mathcal{P}_j(X)A' + \mu_jQ - X_j & A\mathcal{P}_j(X)H'_j\\
				H_j\mathcal{P}_j(X)A' &
				\mu_jR +H_j\mathcal{P}_j(X)H'_j
			\end{array}\right],\\
			\mathcal{R}_j(X)&:=\mu_jR +H_j\mathcal{P}_j(X)H'_j,\ \ \ j\in\mathcal{N}.
		\end{align}
		Then define the following set}
	\begin{align}
		{\Omega := \left\{X\in\mathbb{H}^{n*}|\mathcal{X}(X)\ge 0, \mathcal{R}(X)> 0\right\}.}
	\end{align}
	
	\begin{definition}
		{A solution $Y^+= (Y_1^+,\ldots,Y_N^+)$ to the {CAREs} (\ref{CARE}) is said to be the maximal solution if $Y^+\ge Y$ for any $Y=(Y_1,\ldots,Y_N)$ with $Y_j = \mathcal{P}_j(X)$, $j\in\mathcal{N}$, $X\in\Omega$.}
	\end{definition}
	
	The maximal solution can be numerically computed by solving the following convex programming problem \cite{costa1999maximal}:
	\begin{align}\label{convex}
		\max \left\{{\rm
			tr}\left(\sum_{j=1}^{N}X_j\right):\ \
		X = (X_1,\ldots,X_{N})
		\in \Omega\right\}.
	\end{align}
	
	\begin{lemma}[\cite{costa1999maximal}]\label{le-max}
		Recall Definitions \ref{def-det}
		and \ref{def-stasolution}.
		\begin{itemize}
			\item [(i)] Suppose that $(\mathbf H, \mathbf A,\mathbf p)$ is MS detectable. Then there exists the  maximal solution $Y^+ = (Y_1^+,\ldots,Y_N^+)\in\mathbb{H}_+^n$ to the {CAREs} (\ref{CARE}). Moreover, with $L_i = -K_i(Y^+)$
			for $1\leq i\leq N$, there holds $\rho(\widetilde{\mathcal{L}}^*)\le1$.
			\item [(ii)] There exists at most one MS stabilizing solution to the {CAREs} (\ref{CARE}), which coincides with the maximal
			solution.
		\end{itemize}
	\end{lemma}

	Having the solution $X^+ = (X_1^+,\ldots,X_N^+)$ for the convex programming problem (\ref{convex}), the maximal solution is $Y^+ = (Y_1^+,\ldots,Y_N^+)$ with $Y_j^+ = \mathcal{P}_j(X^+)$. By Lemma  \ref{le-max} (ii), $Y^+$ is also the MS stabilizing solution to the {CAREs} (\ref{CARE}), if the latter exists. The following result is important, which can be applied directly to the main result in this section.
	
	\begin{theorem}\label{th-iff}
		The {CAREs} in (\ref{CARE}) admit the MS stabilizing solution, if and only if
		\begin{itemize}
			
			\item[1)] $(\mathbf H, \mathbf A,\mathbf p)$
			is MS detectable.
			
			\item[2)] $\lambda = 1$ is not an uncontrollable eigenvalue for $(\mathbf A, \mathbf Q)$.
		\end{itemize}
	\end{theorem}
	
	\noindent\textbf{Proof.} 
	See Appendix.	\hfill\rule{2mm}{2mm}
	
	We remark that a similar result to Theorem \ref{th-iff} for the control CAREs is given in Corollary 14 of \cite{ungureanu2015stabilizing}. Nonetheless, a self-contained and independent proof for the {CAREs} (\ref{CARE}) is provided in Appendix, in which the equalities (\ref{adjointinner}) in terms of inner product play an important role. Before presenting the main result in
	this section, we need the following
	technical lemma.
	
	\begin{lemma}\label{le-decom}
		Let $\lambda>0$. If $A'XA = \lambda
		X$ and $QX= 0$ admit a solution $X
		\in\mathbb{S}_+^{n}\backslash\{0\}$, then
		there exist a nonzero vector
		$x_0\in\mathbb{C}^n$ and
		$\omega_0\in\mathbb{R}$
		such that
		\begin{align}\label{Ax0}
			A'x_0 = \sqrt{\lambda}
			e^{j\omega_0}x_0,\ \ Qx_0=0.
		\end{align}
		That is, $\sqrt{\lambda}e^{j\omega_0}$ is
		an uncontrollable eigenvalue for $(A,Q)$ according to the well-known Popov--Belevitch--Hautus (PBH) test.
	\end{lemma}
	
	\noindent\textbf{Proof.}
	The hypothesis on $X\in
	\mathbb{S}_+^{n}\backslash\{0\}$ implies
	that $X=GG'$ with $G\in
	\mathbb{R}^{n\times r}$ and
	$r={\rm rank}\{X\}>0$. It follows
	that $\lambda GG'=A'GG'A$ and
	$QGG'=0$. Hence, there exists an
	orthogonal matrix $U\in
	\mathbb{R}^{r\times r}$ such that
	\begin{align}\label{GU}
		\sqrt{\lambda}GU=A'G,\ \ \ QGU=0.
	\end{align}
	Since all the eigenvalues of
	an orthogonal matrix are on the
	unit circle, there exists a
	nonzero vector $u_0\in \mathbb{C}^r$
	such that $Uu_0=e^{j\omega_0}u_0$
	for some $\omega_0\in\mathbb{R}$.
	Multiplying the two equalities in
	(\ref{GU}) by $u_0$ from right
	verifies the eigenvalue--eigenvector equation and $Qx_0=0$ in
	(\ref{Ax0}) by taking $x_0=Gu_0$.
	\hfill\rule{2mm}{2mm}

	\begin{theorem}\label{th-iffMar}
		The {CAREs} in (\ref{CARE}) admit
		the MS stabilizing solution, if
		and only if
		\begin{itemize}
			
			\item[1)]
			$(\mathbf H, \mathbf A,\mathbf p)$ is MS detectable;
			
			\item[2)]
			{$(A, Q)$}
			does not have uncontrollable eigenvalues on the unit circle, i.e.,
			\begin{align}\label{rank}
				{\rm rank}\left\{\left[\begin{array}{cc}
					\lambda I_{n} -A & Q
				\end{array}\right]
				\right\} = n, \ \forall |\lambda|=1,
			\end{align}
			where $\lambda$ is an eigenvalue of $A$.
		\end{itemize}
	\end{theorem}
	
	\noindent\textbf{Proof.}
	By Theorem \ref{th-iff}, it suffices to show that $1$ is an uncontrollable eigenvalue for $(\mathbf A,\mathbf Q)$,
	if and only if there exists
	some eigenvalue on the unit circle
	that is uncontrollable for $(A, Q)$.
	We first show the sufficiency. By the
	PBH test, there exist some $|\lambda|=1$
	and $0\neq v\in\mathbb{C}^n$ such that
	\begin{align}
		A'v = \lambda v,\ \ \ Qv = 0.
	\end{align}
	Let $V_i := vv^* + \bar v\bar v^*\in
	\mathbb{S}_+^{n}$ for all
	$i\in\mathcal{N}$ such that $V = (V_1,\ldots V_N)\in{\mathbb H}_+^n$.
	Then by (\ref{Listar}),
	\begin{align}
		\mathcal{L}_i^*(V) =&
		\sum_{j=1}^{N}p_{ij}A'V_iA\notag
		=A'(vv^* +\bar v\bar v^*)A\notag\\
		=&A'v(A'v)^* + A'\bar v(A'\bar
		v)^*\notag\\
		=&\lambda\bar\lambda vv^* +
		\lambda\bar\lambda
		\bar v\bar v^* = V_i\ \
		\forall\ i\in\mathcal{N}.
	\end{align}
	In addition, it is easy to see
	that $QV_i =0$ $\forall i\in
	\mathcal{N}$. Therefore, 1 is an uncontrollable eigenvalue for
	the pair $(\mathbf A,\mathbf Q)$.
	
	To show the necessity, assume
	that there exists $V=(V_1,\ldots,V_N) \in{\mathbb H}_+^n\backslash\{0\}$ such that  $\mathcal{L}_i^*(V) = V_i$ and $QV_i=0$. Recall the limit probability
	distribution $\mu$ in (\ref{tpmmu})
	of Lemma \ref{le-tpm} where $\mu=
	\left[\begin{array}{cccc}
		\mu_1 & \mu_2 & \cdots & \mu_N
	\end{array}\right]\neq 0$
	is a positive row vector.
	Since $\mu P=\mu$, there holds
	\begin{align*} 
		\mu_j = \sum_{i=1}^N \mu_ip_{ij}\
		\ \ \forall\ j\in \mathcal{N}.
	\end{align*}
	By the above equality,
	$V_i=\mathcal{L}_i^*(V)$ with
	$\mathcal{L}_i^*(V)$ defined in
	(\ref{Listar}), and $QV_i=0$
	$\forall i\in\mathcal{N}$, we have
	\begin{align*}
		X:=& \sum_{i=1}^N \mu_iV_i =
		\sum_{i=1}^N \mu_i\mathcal{L}_i^*(V)
		=  \sum_{i=1}^N \sum_{j=1}^N
		\mu_ip_{ij}A'V_jA \\
		=& \sum_{j=1}^N A'V_jA
		\left(\sum_{i=1}^N\mu_ip_{ij}\right)
		= \sum_{j=1}^N \mu_jA'V_jA \\
		=& A'\left(\sum_{j=1}^N
		\mu_jV_j\right)A=A'XA,\\
		QX =& \sum_{i=1}^N\mu_iQV_i = 0.
	\end{align*}
	There thus hold $A'XA = X$
	and $QX=0$. An application of
	Lemma \ref{le-decom} with $\lambda=1$
	concludes that $(A,Q)$ has at least
	an uncontrollable eigenvalue on the
	unit circle. The proof is complete. \hfill\rule{2mm}{2mm}
	
	Condition 2) in Theorem \ref{th-iffMar}
	shows that the controllability requirement
	is the same as that arising in the study
	of the standard ARE \cite{gu2012discrete},
	which is an interesting result. In
	Theorem \ref{th-iffMar}, the matrix
	rank condition (\ref{rank}) is simple
	to check, while the MS detectability
	for $(\mathbf H, \mathbf A,\mathbf p)$
	is not straightforward, which needs to
	be numerically verified by solving a feasibility problem generally, in
	terms of $2^m$ linear matrix inequalities
	(LMIs); see (\ref{lmis}) in the next section. The other issue is the exponential complexity of the optimal stationary estimator in Theorem \ref{th-estimator}. In the next
	section, we will derive some sufficient
	and necessary conditions for the MS detectability by exploring the system
	structure, which show directly how
	system parameters influence the MS detectability, and propose a locally optimal stationary estimator that has a linear complexity.
	
	\section{MS Detectability and Locally Optimal Stationary Estimator}\label{sec-det}
	\subsection{MS detectability for  $(\mathbf H, \mathbf A,\mathbf p)$}\label{subsec-det}
	
	Theorem 3.9 of \cite{costa2005discrete} describes the MS stability of MJLSs, based on which we also have the following definition
	for the MS detectability for system (\ref{jumpxyk}), coinciding exactly with Definition \ref{def-det}.
	
	\begin{definition}\label{def-detsingle}
		We say that $(\mathbf H, \mathbf A,\mathbf p)$ is MS detectable, if there exist $\{X_i>0\}_{i=1}^{N}$ and $L=(L_1,\ldots,L_{N})\in\mathbb{H}^{n,m}$  such that either of the following two inequalities holds:
		\begin{align}\label{detect1} 
			X_i &> \sum_{j=1}^{N}
			p_{ij}(A+L_jH_j)'X_j(A+L_jH_j)\
			\ \forall i\in\mathcal{N},
			\\ \label{detect2} 
			X_j &> \sum_{i=1}^{{N}}
			p_{ij}(A+L_iH_i)X_i(A+L_iH_i)'\ \
			\forall j\in\mathcal{N}.
		\end{align}
	\end{definition}
	
	{The pair $(C,A)$ is assumed to be detectable throughout the section, which is clearly weaker than the MS detectability of	$(\mathbf H, \mathbf A,\mathbf p)$ that relates to channel parameters of packet drops.}
	We first provide an analytic
	necessary condition for the MS detectability.
	
	
	\begin{theorem}\label{th-nece}
		The triple
		$(\mathbf H, \mathbf A,\mathbf p)$
		is MS detectable, only if
		\begin{align}\label{nece}
			\prod_{i=1}^{N}(1-q_i)\rho^2(A)<1.
		\end{align}
	\end{theorem}
	
	\noindent\textbf{Proof.}
	By (\ref{detect2}),
	the MS detectability for $(\mathbf H, \mathbf A,\mathbf p)$ implies that there exist $\{X_i>0\}_{i=1}^{N}$ such that
	\begin{align}
		X_1 &>\sum_{i=2}^{{N}}p_{ij}
		(A+L_iH_i)X_i(A+L_iH_i)'\\ \nonumber
		& \ \ \ +
		\prod_{i=1}^{N}(1-q_i)AX_1A
		\ge\prod_{i=1}^{N}(1-q_i)AX_1A'.
	\end{align}
	This implies that
	$A\sqrt{\prod_{i=1}^{N}(1-q_i)}$ must be
	a Schur stability matrix, leading to
	the inequality in (\ref{nece}). \hfill\rule{2mm}{2mm}
	
	The MS detectability for
	$(\mathbf H, \mathbf A,\mathbf p)$
	can be numerically verified by solving
	the following feasibility problem of
	LMIs that can be obtained by using the
	MS detectability condition in
	(\ref{detect1}) and the Schur complement
	repeatedly.
	
	
	\begin{proposition}\label{prop}
		Denote $X_{\rm d}:={\rm diag}\{X_1,\ldots,X_N\}$ and
		$\Psi_i(X,\Omega) = \left[\ 
		\Psi_{i1}\ \Psi_{i2}\ \cdots\
		\Psi_{iN}\ \right]$ where
		$\Psi_{ij}=\sqrt{p_{ij}}(A'X_j+H_j'\Omega_j)$
		for $1\leq j\leq N$.
		Then, $(\mathbf H, \mathbf A,\mathbf p)$
		is MS detectable, if and only if there
		exist $\{X_i>0\}_{i=1}^{N}$ and $\{\Omega_i
		\in\mathbb{R}^{n\times m}\}_{i=1}^{N}$
		such that LMI 
		\begin{align}\label{lmis}
			\left[\begin{array}{cc}
				X_i & \Psi_i(X,\Omega)\\
				\Psi_i(X,\Omega)' & X_{\rm d}
			\end{array}\right]>0
		\end{align}
		holds for each $i\in\mathcal{N}$.
	\end{proposition}
	
	A significant challenge to checking the MS detectability  from Proposition \ref{prop} lies in the
	complexity, due to $N=2^m$ LMIs in (\ref{lmis}).
	We will {focus} on lowering the complexity
	from $N=2^m$ to $2m$. This is indeed
	possible through decoupling the MS
	detectability for $(\mathbf H, \mathbf A,\mathbf p)$ into that for $m$
	subsystems, respectively.
	
	Denote the state estimation error by
	\begin{align}
		e(k) := x(k) - \hat{x}(k).
	\end{align}
	Taking the difference between (\ref{xk}) and (\ref{hatx}) yields
	\begin{align}\label{ek}
		e(k+1) = (A - L_{\theta(k)}H_{\theta(k)})e(k)
	\end{align}
	after removing the terms about noises.
	Then the existence
	of $\{L_{\theta(k)}\}$ that achieve
	the MS stability for the error dynamics
	described in (\ref{ek}) is equivalent
	to the MS detectability for
	$(\mathbf H,\mathbf A,\mathbf p)$.
	Without loss of generality, the pair
	$(C,A)$ for system (\ref{xyk}) is
	assumed to be of the following
	Wonham decomposition form
	\cite{wonham1967pole}:
	\begin{align}\label{wondecom}
		A = \left[\begin{array}{cccc}
			A_1 & 0 & \cdots & 0\\
			A_{21} & A_2 & \ddots & \vdots\\
			\vdots & \ddots & \ddots & 0\\
			A_{m1} & \cdots & A_{m(m-1)} & A_m
		\end{array}\right],\  C = \left[\begin{array}{cccc}
			c_1 & 0 & \cdots &0\\
			0 & c_2 & \ddots & \vdots\\
			\vdots & \ddots & \ddots & 0\\
			0 & \cdots & 0 & c_m
		\end{array}\right],
	\end{align}
	where  $A_i\in\mathbb{R}^{n_i\times
		n_i}$, $c_i\in\mathbb{R}^{1\times n_i}$,
	$\sum_{i=1}^{m}n_i=n$, and each pair
	$(c_i, A_i)$ is detectable under the
	detectability of $(C,A)$. Define
	\begin{align*}
		\theta_i(k)&:=\gamma_i(k)+1,\ \ \
		h_{i,\theta_i(k)}:=\gamma_i(k)c_i, \\
		\mathbf A_i &:= (A_i,A_i),\
		\mathbf h_i := (h_{i,1}, h_{i,2}),\ \mathbf p_i := \{p_{lj}^{(i)}\},
	\end{align*}
	for $1\leq i\leq m$, where $p_{lj}^{(i)}$
	is the $(l,j)$th element of the TPM $P_i$ in (\ref{tpm}). Let us introduce a particular $L_{\theta(k)}$ that is in the
	block diagonal form, conformal to that
	of $C$:
	\begin{align}\label{diagL}
		\bar L_{\theta(k)} := {\rm
			diag}\{\ell_{\theta_1(k)},\ldots,
		\ell_{\theta_m(k)}\},\ \
		\ell_{\theta_i(k)}
		\in\mathbb{R}^{n_i}.
	\end{align}
	
	\begin{theorem}\label{th-decouple}
		If $(\mathbf h_i,\mathbf A_i,\mathbf p_i)$ is MS detectable for all $1\leq i\leq m$,
		then $(\mathbf H,\mathbf A,\mathbf p)$ is MS detectable.
	\end{theorem}
	
	\noindent\textbf{Proof.}
	For system (\ref{jumpxyk}), consider a
	similarity transform $x(k) =
	S\tilde x(k)$ with
	\begin{align}
		{S = {\rm diag}\{I_{n_1},\epsilon^{-1}
			I_{n_2},\ldots, \epsilon^{1-m} I_{n_m}\}},\ \ \epsilon>0.
	\end{align}
	Then the MS detectability for
	$(\mathbf H,\mathbf A,\mathbf p)$, i.e.,
	system (\ref{jumpxyk}), is equivalent to that for
	\begin{subequations}
		\begin{align}
			\tilde x(k+1) &= \tilde A\tilde x(k)
			+ S^{-1}w(k),\\
			{y_{\rm r}(k)} &= \tilde
			H_{\theta(k)}\tilde x(k) + D_{\theta(k)}v(k),
		\end{align}
	\end{subequations}
	where $\tilde A = S^{-1}AS$ and
	$\tilde H_{\theta(k)} = H_{\theta(k)}S
	=\Gamma_{\theta(k)}CS$ that is in the
	block diagonal form.
	Define $\tilde L_{\theta(k)} := S^{-1}\bar L_{\theta(k)}$, where $\bar L_{\theta(k)}$ is given by (\ref{diagL}). It follows from (\ref{ek}) that
	\begin{align}\label{tildeek}
		\tilde e(k+1) = (\tilde A -
		\tilde L_{\theta(k)}\tilde H_{\theta(k)})\tilde e(k),
	\end{align}
	where $\tilde e(k) = S^{-1}e(k)$.
	Specifically, $\tilde A - \tilde L_{\theta(k)}\tilde H_{\theta(k)}$
	has the following lower triangular form
	\begin{align*}
		&\tilde A - \tilde L_{\theta(k)}\tilde
		H_{\theta(k)}
		{=S^{-1}(A-\bar L_{\theta(k)}
			H_{\theta(k)})S}\notag\\
		=& \left[\begin{array}{cccc}
			\mathcal{A}_{\theta_1(k)} & 0
			& \cdots & 0\\
			{\epsilon A_{21}} & \mathcal{A}_{\theta_2(k)}
			& \ddots & \vdots\\
			\vdots & \ddots & \ddots & 0\\
			{\epsilon^{m-1}A_{m1}} & \cdots & {\epsilon A_{m(m-1)}} &
			\mathcal{A}_{\theta_m(k)}
		\end{array}\right],
	\end{align*}
	where
	$\mathcal{A}_{\theta_i(k)} = A_i-\ell_{\theta_i(k)}
	h_{i,\theta_i(k)}$. Hence as $\epsilon\to 0$, estimation error dynamics (\ref{tildeek}) approaches to a diagonal form, implying that with the block diagonal gain $\bar L_{\theta(k)}$ in (\ref{diagL}), the MS detectability for $\{\mathbf h_i,\mathbf A_i,\mathbf
	p_i\}_{i=1}^m$ is equivalent to the MS
	stability for the error dynamics in (\ref{tildeek}). The proof is thus complete.	\hfill\rule{2mm}{2mm}

	\begin{remark}
		The Wonham decomposition (\ref{wondecom}), which gives $m$ detectable subsystems $(c_i,A_i)$ from the original detectable system $(C,A)$, plays a critical role in lowering
		the complexity from $N=2^m$ to
		$2m$, without which the complexity
		reduction is not possible. This decomposition is also a powerful tool when dealing with networked stabilization for multi-input systems over logarithmic quantization and i.i.d. fading channels \cite{gu2008networked,gu2015networked,xiao2012feedback}. \pfbox
	\end{remark}
	Now we obtain a
	sufficient condition on decoupling
	the MS detectability, but the complexity
	of the stationary optimal estimator
	remains exponential. In the next subsection, we propose a locally optimal stationary estimator, which has the complexity $2m$ as well.
	
	\subsection{A locally optimal stationary estimator}
	
	In this subsection, instead of the 
	general estimator gain $L_{\theta(k)}$ in estimator (\ref{hatx}), we consider the block diagonal gain in (\ref{diagL}) that reduces estimator (\ref{hatx}) to 
	\begin{align}\label{subhatx}
		\hat x(k+1) = A\hat x(k) +
		\bar L_{\theta(k)}[y_{\rm r}(k) -
		H_{\theta(k)}\hat x(k)].
	\end{align}
	Write system state $x(k)$ as $x(k)=[x_1(k)',\ldots,x_m(k)']'$ with $x_i(k)\in\mathbb{R}^{n_i}$ and measurement $y_{\rm r}(k)$ as $y_{\rm r}(k)=[y_{{\rm r},1}(k),\ldots,y_{{\rm r},m}(k)]'$, and let $\hat x_i(k)$ be the state estimation of $x_i(k)$. It is observed that  estimator (\ref{subhatx}) can be written as the following $m$ sub-estimators:
	\begin{numcases}{}\label{subhatxi}
		\mspace{-6mu} \hat x_1(k+1) = A_1\hat x_1(k) +
		\ell_{\theta_1(k)}[y_{{\rm r},1}(k) -
		h_{1,\theta_1(k)}\hat x_1(k)],\notag\\
		\mspace{-6mu} \hat x_i(k+1) = A_i\hat x_i(k) +
		\ell_{\theta_i(k)}[y_{{\rm r},i}(k) -
		h_{i,\theta_i(k)}\hat x_i(k)] \notag\\
		~~~~~~~~~~~~~~~~~~ + \sum_{j=1}^{i-1}A_{ij}\hat x_j, \ 2\le i \le m.
	\end{numcases}
	Let $\{Q_i\}_{i=1}^m$ and $\{R_i\}_{i=1}^m$ be the sub-matrices on the diagonal of $Q$ and $R$, respectively. We have the following estimation result that is similar to that in Theorem \ref{th-estimator}. Hence, its proof is omitted.
	\begin{lemma}\label{le-subestimator}
		Assume that there exists the MS stabilizing solutions
		$\{Z_i = (Z_{i,1},Z_{i,2})\}$ to the following {CAREs}
		\begin{align}\label{CAREi}
			Z_{i,r} = & \sum_{j=1}^{2}p_{jr}^{(i)}
			\big\{A_jZ_{i,j}[{I_{n_i}}+h_{i,j}'(\pi_{i,j}R_i)^{-1}
			h_{i,j}Z_{i,j}]^{-1}A_j' \notag\\
			&~~~~~~~+ \pi_{i,j}Q_j\big\},\ \
			\ 1\leq i\leq m,\  1\leq r\leq 2.
		\end{align}
		Then for each sub-estimator in (\ref{subhatxi}), the optimal stationary gains that minimize
		\begin{align}\label{costi}
			J_i(\infty) = \lim_{k\to
				\infty}\mathrm{E}[\|x_i(k)
			- \hat x_i(k)\|^2]
		\end{align}
		for $i=1,\ldots,m$
		are given by
		\begin{align}
			\ell_{\theta_i(k)} = K_i(Z_{i}) := A_iZ_{i,2}c_i'(c_iZ_{i,2}c_i'
			+ \pi_{i,2}R_i)^{-1}
		\end{align}
		for both $\theta_i(k) = 1,2$.
	\end{lemma}

	Note that we have intentionally allowed the value of $l_{\theta_i(k)}$ for the case $\theta_i(k) = 1$  to be the same as that for $\theta_i(k) = 2$, without any effect, such that only a gain is required for each sub-estimator.
	Define $\Sigma(k) := \mathbb{E}[e(k)e(k)']$ and $\Sigma_i(k) := \mathbb{E}[e(k)e(k)'\mathbf{1}_{\theta(k)=i}]$. Notice that
	\begin{align}
		\Sigma(k) = \sum_{i=1}^{N}\Sigma_i(k).
	\end{align}
	In light of Lemma \ref{le-subestimator}, we present the following locally optimal estimator.
	
	\begin{proposition}\label{pro-sub}
		For the system dynamics described in (\ref{jumpxyk}) with $(C,A)$ in the Wonham decomposition form (\ref{wondecom}), a locally optimal stationary estimator of which sub-estimator $i$ in (\ref{subhatxi}) minimizes cost $J_i(\infty)$
		in (\ref{costi}) is given by
		\begin{align}\label{loshatx}
			\hat x(k+1) = A\hat x(k) +
			\bar K[y_{\rm r}(k) -
			H_{\theta(k)}\hat x(k)],
		\end{align}
		where $\bar K = {\rm diag}\{K_1(Z_1),\ldots,K_m(Z_m)\}$. Moreover, the corresponding stationary error covariance $\Sigma(\infty)$ is given by the solution of following coupled Lyapunov equations
		\begin{align}
			\Sigma_j(\infty) =& \sum_{i=1}^{N}p_{ij}
			\big\{[A-\bar KH_i]
			\Sigma_i(\infty)[A-\bar KH_i]'\notag\\
			+&\mu_i[\bar KD_iRD_i'\bar K'+Q]
			\big\}, \ j\in\mathcal{N}.
		\end{align}
	\end{proposition}
	
	\begin{remark}
		Compared with the complexity of solving $2^m$ CAREs for the optimal stationary estimator, the estimator in (\ref{loshatx}) only needs to solve $2m$ CAREs and one estimator gain $\bar K$. {This is achieved by restricting the estimator gain to the block diagonal form (\ref{diagL}) and the estimation cost to local costs (\ref{costi}). As it is expected, the estimator in (\ref{loshatx}) has a performance loss compared with the optimal one, which will be illustrated in the simulation.} Nonetheless, it follows from Theorems \ref{th-iffMar} and \ref{th-decouple} that if $(\mathbf h_i,\mathbf A_i,\mathbf p_i)$ is MS detectable for $1\leq i\leq m$ and the rank condition (\ref{rank}) for each $(A_i, Q_i)$ holds, the error dynamics described in (\ref{ek}) with $L_{\theta(k)} = \bar K$ is MS stable, implying that the covariance of the locally optimal stationary estimator (\ref{loshatx}) converges to a finite value $\Sigma(\infty)$. \pfbox
	\end{remark}
	
	\begin{remark}
		{	
			If $m>n$, it is not possible to obtain the Wonham decomposition form in (\ref{wondecom}), but in the following form:
			\begin{align*}
				A = \left[\begin{array}{cccc}
					A_1 & 0 & \cdots & 0\\
					A_{21} & A_2 & \ddots & \vdots\\
					\vdots & \ddots & \ddots & 0\\
					A_{n1} & \cdots & A_{n(n-1)} & A_{n}
				\end{array}\right],\  C = \left[\begin{array}{cccc}
					c_1 & 0 & \cdots &0\\
					0 & c_2 & \ddots & \vdots\\
					\vdots & \ddots & \ddots & 0\\
					0 & \cdots & 0 & c_{n}\\
					& & C_0&
				\end{array}\right],
			\end{align*}
			where $C_0=[C_{n+1}',\ldots, C_m']'T$ with $T$ the Wonham decomposition transform matrix. In this case, we may not obtain the ideal locally optimal estimator (\ref{loshatx}) that admits the low complexity $2m$ and uses measurement information from all $m$ sensors. One simple way is to only use the measurements from the $n$ sensors and abandon the left $m-n$ sensors. One can also expect a better method to fuse the measurements from both the $n$ and $m-n$ sensors, which deserves a further study in the future.}\pfbox
	\end{remark}

	Since the MS detectability for each subsystem $(\mathbf h_i, \mathbf A_i, \mathbf p_i)$ is the sufficient MS detectability condition for $(\mathbf H, \mathbf A, \mathbf p)$, and is also required by the locally optimal stationary estimator, we will investigate the MS detectability for $(\mathbf h_i, \mathbf A_i, \mathbf p_i)$ by providing some analytic MS detectability conditions in the
	next subsection.
	
	\subsection{MS detectability for $(\mathbf h_i, \mathbf A_i, \mathbf p_i)$}
	
	For convenience, we will omit the subscript $i$ in this subsection, i.e., denote $A := A_i$, $C:=c_i$, $\gamma(k) := \gamma_i(k)$, $\theta(k) := 1+\gamma(k)$ for any $i=1,\ldots,m$ with a slight abuse of notation. Also, the TPM for the process of Markovian packet drops $\{\gamma(k)\}$ is now given by
	\begin{align}
		P  = \left[\ p_{ij}\ \right] =
		\left[\begin{array}{cc}
			1-q & q\\ p & 1-p
		\end{array}\right].
	\end{align}
	We further denote $C_{\theta(k)} := \theta(k)C$ and
	\begin{align*}
		\mathbf A := (A,A),\ \
		\mathbf{C} := (C_1,C_2),\ \
		\mathbf p := \{p_{ij}\},\ \ i,j\in\{1,2\}.
	\end{align*}
	For simplicity, define the following operators
	\begin{subequations}
		\begin{align}
			\phi_1(L,X_1,X_2) &:= (1-q)A'X_1A +
			qA_L'X_2A_L,\\
			\phi_2(L,X_1,X_2) &:= pA'X_1A +
			(1-p)A_L'X_2A_L,\\
			\psi_1(L,X_1,X_2) &:= (1-q)AX_1A' +
			pA_LX_2A_L',\\
			\psi_2(L,X_1,X_2) &:= qAX_1A' +
			(1-p)A_LX_2A_L',\\
			g_1(X_1,X_2)&:=(1-q)AX_1A'+
			pAX_2A' \\ \notag &~~~~~~ -p AX_2C'(CX_2C')^{-1}CX_2A',\\
			g_2(X_1,X_2)&:=qAX_1A'+ (1-p)AX_2A'
			\\ \notag 	&~~~~~~ -(1-p)
			AX_2C'(CX_2C')^{-1}CX_2A',
		\end{align}
	\end{subequations}
	where $A_L = A + LC$.
	The next lemma is useful.
	
	\begin{lemma}\label{le-eq}
		The following statements are equivalent.
		\begin{itemize}
			\item [a)] The triple
			$(\mathbf p,\mathbf C,\mathbf A)$ is MS detectable.
			\item [b)] There exist $X_1>0, X_2>
			0$, and $L\in\mathbb{R}^{n_i\times m}$
			such that $ X_i>\psi_i(L,X_1,X_2)$
			for $i=1,2$.
			\item [c)] There exist $X_1>0$
			and $X_2> 0$ such that $X_i>
			g_i(X_1,X_2)$ for $i=1,2$.
			\item[d)] There exist $X_1>0, X_2>0$,
			and $\Omega\in\mathbb{R}^{n_i\times m}$
			such that the following LMIs hold:
			\begin{align*}
				\left[\begin{array}{ccc}
					X_1 & \sqrt{1-q}A'X_1 &
					\sqrt{q}(A'X_2+ C'\Omega')\\
					* & X_1 & 0\\
					* & 0 & X_2
				\end{array}\right]&>0,\notag\\
				\left[\begin{array}{ccc}
					X_2 & \sqrt{p}A'X_1 &
					\sqrt{1-p}(A'X_2+ C'\Omega')\\
					* & X_1 & 0\\
					* & 0 & X_2
				\end{array}\right]&>0.
			\end{align*}
		\end{itemize}
	\end{lemma}
	
	\noindent\textbf{Proof.}
	a) $\Leftrightarrow$ b): This is straightforward in accordance with
	Definition \ref{def-detsingle}.
	
	c) $\Rightarrow$ b): For any $L\in
	\mathbb{R}^{n_i\times m}$ and $X>0$,
	\begin{align*}
		(A+LC)X(A+LC)'
		=&AXA'-  AXC'(CXC')^{-1}CXA' \\
		& \ \ \ + (L+L_X)CXC'(L+L_X)',
	\end{align*}
	where $L_X=AXC'(CXC')^{-1}$. As a result,
	we have
	\begin{align*}
		\psi_1(-L_{X_2},X_1,X_2) &=
		g_1(X_1,X_2), \\
		\psi_2(-L_{X_2},X_1,X_2) &= g_2(X_1,X_2).
	\end{align*}
	The above two equalities imply
	\begin{align}
		\psi_i(L,X_1,X_2) \ge g_i(X_1,X_2),\
		\ i=1,2,
	\end{align}
	for any $X_1>0$, $X_2>0$, and
	$L\in \mathbb{R}^{n_i\times m}$.
	Therefore, if $ X_1>g_1(X_1,X_2)$ and
	$X_2>g_2(X_1,X_2)$ hold for some
	$X_1>0$ and $X_2>0$, then $ X_1>
	\psi_1(-L_{X_2},X_1,X_2)$ and
	$X_2>\psi_2(-L_{X_2},X_1,X_2)$,
	proving the statement of c)
	$\Rightarrow$ b).
	
	b) $\Rightarrow$ c): This is clearly true
	by $X_1> \psi_1(L,X_1,X_2)\ge g_1(X_1,X_2)$
	and $X_2>\psi_2(L,X_1,X_2)\ge
	g_2(X_1,X_2)$.
	
	a) $\Leftrightarrow$ d): This is straightforward from Proposition \ref{prop}. \hfill\rule{2mm}{2mm}
	
	\begin{theorem}
		The triple $(\mathbf p,\mathbf C,\mathbf A)$ is MS detectable, if
		\begin{align}\label{suff}
			\min\{q,1-p\}>\lambda_{\rm c}=
			1-\frac{1}{\prod_{i}\max
				\{|\lambda_i(A)|^2,1\}},
		\end{align}	
		where $\lambda_i(A)$ is the $i$th
		eigenvalue of $A$.
	\end{theorem}
	
	\noindent\textbf{Proof.}
	Recall the hypothesis on the
	detectability of $(C,A)$, assumed
	throughout this section. Hence,
	$(c_i,A_i)$ is detectable with
	${\rm rank}\{c_i\}=1$ for
	$1\leq i\leq m$. Then by Lemma 5.4 of \cite{schenato2007foundations}, there exists $X> 0$ such that
	\begin{align}\label{gammac}
		X > AXA'-  \lambda AXC'(CXC')^{-1}CXA',
	\end{align}
	if and only if $\lambda>\lambda_{\rm c}$.
	Therefore, if (\ref{suff}) holds, there
	exists $\bar X> 0$  to (\ref{gammac}) with
	$\lambda = \min\{q,1-p\}$ such that
	\begin{align*} 
		\bar X &> A\bar XA'-  q A\bar
		XC'(C\bar XC')^{-1}C\bar
		XA',\\ 
		\bar X &> A\bar XA'- (1-p)A\bar
		XC'(C\bar XC')^{-1}C\bar XA'.
	\end{align*}
	Let $\bar X = X_1 = pX_2/q$. Then,
	the above inequalities reduce to
	\begin{align*}
		X_1 >g_1(X_1,X_2),\ \ \
		X_2>g_2(X_1,X_2).
	\end{align*}
	Therefore, the sufficient condition in (\ref{suff}) holds, following from the equivalence of a) and c) in Lemma \ref{le-eq}. \hfill\rule{2mm}{2mm}
	
	When $m=n$, i.e., the order of $A_i$ is
	one for $1\leq i\leq n$, we have the
	following analytic necessary and sufficient condition for the MS detectability of $(\mathbf p,\mathbf C,\mathbf A)$.
	
	\begin{theorem}\label{th-anaiff}
		If $m=n$, then the triple
		$(\mathbf p,\mathbf C,\mathbf A)$
		is MS detectable, if and only if
		\begin{align}\label{iffq}
			q>1-\frac{1}{\rho(A)^2}.
		\end{align}
	\end{theorem}
	
	\noindent\textbf{Proof.}
	The argument for necessity is the same as that in Theorem \ref{th-nece}. So we
	only show the sufficiency. First, we observe that if there exists a matrix $L$ such that $A_L=A+LC=0$, then the MS detectability
	condition in (\ref{detect2}) becomes the
	case that there exist $X_1>0, X_2>0$ such that
	\begin{align}\label{ineAL0}
		X_1>(1-q)AX_1A', \ X_2>pAX_1A'.
	\end{align}
	In this case, if $q>1-\rho(A)^{-2}$, we
	can always find some $X_1>0$ and
	$X_2>0$, rendering inequalities in (\ref{ineAL0}) true. Clearly, when
	$A$ and $C$ are both scalars, the
	choice of $L = -A/C$ makes $A_L=0$,
	which completes the proof. \hfill\rule{2mm}{2mm}
	
	\begin{remark}
		It is worth mentioning that the results in Section \ref{sec-existence} and this subsection
		can be applied to the dual optimal control
		problem studied in \cite{mo2013lqg}, of which focus is on the convergence issue under assumptions of stabilizability and detectability of the system.
		To be specific, consider the 
		linear system described by
		\begin{align}
			x({k+1)}=Ax(k)+Bu_{\rm r}(k),\
			\ u_{\rm r}(k) = \gamma(k)u(k),
		\end{align}
		where $u(k)\in\mathbb{R}^m$ is the control
		signal sent from the remote controller via
		the Markovian packet drop channel. In
		accordance with the MJLS theory in
		\cite{costa2005discrete}, the optimal
		controller is given by $u(k) = F(X)x(k)$,
		where $F(X)$ is computed using the
		MS stabilizing solution to following
		control {CAREs}:
		\begin{align}\label{controlCARE}
			X_i &= A'\mathcal{E}_i(X)A + W +
			A'\mathcal{E}_i(X)B_iF_i(X),\\
			F(X) &= -(U +
			B'\mathcal{E}_i(X)B)^{-1}B'
			\mathcal{E}_i(X)A, \ i = 1,2.\notag
		\end{align}
		Here, $\mathcal{E}_i(X) = \sum_{j=1}^{2}p_{ij}X_j$, $B_1 = 0, B_2 = B$ and
		$(W\ge 0, U>0)$ are the weighting
		matrices for the
		state and control signal, respectively.
		According to the notion of MS stabilizability
		in Definition 2 of \cite{costa1995discrete}, which is dual to the MS detectability, similar results can be obtained for such an optimal control problem.   \pfbox
	\end{remark}
	
	\section{Simulation Examples}\label{sec-examples}
	
	\subsection{Existence of the MS stabilizing solution}
	First, we illustrate the theoretical results about the
	MS stabilizing solution and MS
	detectability by a numerical example. Consider a third-order system of the form in \ref{xyk} with 
	$C = R = I_3$, and
	\begin{align}
		A &= \left[\begin{array}{ccc}
			1 & 0 & 0\\
			1 & 1.2 & 0\\
			1 & 1.5 & 1.3
		\end{array}\right],\ \ Q = \left[\begin{array}{ccc}
			1 & 0 & 0\\
			0 & 1 & 0\\
			0 & 0 & 0
		\end{array}\right].
	\end{align}
	Note that $(C,A)$ is already in the form
	of Wonham decomposition. Clearly, $\lambda=1.3$ is an uncontrollable eigenvalue for $(A, Q)$.
	Nonetheless, according to the condition 2) in Theorem \ref{th-iffMar}, we only require
	that $|\lambda| = 1$ be controllable.
	This condition is satisfied in this
	example. Let the parameters of the three
	channels be
	\begin{align*}
		p_1 = 0.50,\ \ \ p_2 = 0.60,\ \ \
		p_3 = 0.70, \\
		q_1=0.20,\ \ \ q_2=0.32,\ \ \
		q_3 = 0.51.
	\end{align*}
	So, $q_1>0, q_2>1-1.2^{-2} = 0.3056$, and
	$q_3>1-1.3^{-2} = 0.4083$. Then by Theorems
	\ref{th-decouple} and \ref{th-anaiff}, the
	system is MS detectable. Therefore, by
	Theorem \ref{th-iffMar}, the MS stabilizing
	solution to the {CAREs} (\ref{CARE}) exists. By solving the convex programming
	problem (\ref{convex}) using YALMIP \cite{Lofberg2004}, we can obtain the stabilizing solution $Y$. Then with $\{K_i(Y)\}_{i=1}^8$ computed by (\ref{Kj}), we have $\rho(\widetilde{\mathcal{L}}^*)
	= \rho(\mathcal{A}) = 0.9297 <1$,
	according to Remark 3.5 of \cite{costa2005discrete}, where
	$$
	\mathcal{A}=(P'\otimes I_9){\rm diag}
	\{(A-K_i(Y)H_i)\otimes (A-K_i(Y)H_i)\}_{i\in\mathcal{N}}.
	$$
	Thus, $Y$ is indeed the MS stabilizing solution from Definition \ref{def-stasolution}.
	
	\subsection{Performance of the optimal and locally optimal stationary estimators}
	
	We will use a target tracking example \cite{singer1970estimating}
	to show the estimation performance of the optimal and locally optimal stationary estimators in Theorem \ref{th-estimator} and Proposition \ref{pro-sub}, respectively. For brevity, the two estimators will be abbreviated as OS estimator and LOS estimator, respectively. The system
	dynamics is described by \cite{singer1970estimating}
	\begin{align}
		x(k+1) = \left[\begin{array}{ccc}
			1 & 0 & 0\\
			T & 1 & 0\\
			T^2/2 & T & 1
		\end{array}\right]x(k) + w(k),
	\end{align}
	where $T$ is the sampling period and $w(k)$
	is the Gaussian noise with covariance
	\begin{align}
		Q = 2\alpha \sigma_m^2\left[\begin{array}{ccc}
			T & T^2/2 & T^3/6\\
			T^2/2 & T^3/3 & T^4/8\\
			T^3/6 & T^4/8 & T^5/20
		\end{array}\right],
	\end{align}
	with $\alpha$ the reciprocal of the maneuver
	time constant and $\sigma_m^2$ the variance
	of the target acceleration. The first,
	second and third entries of $x(k)$
	represent the acceleration, speed and
	position of the target, respectively.
	Suppose that there are three sensors
	measuring the target acceleration,
	speed and position, respectively. As a result,
	the measurement model is given by
	\begin{align}
		y(k) = \left[\begin{array}{ccc}
			1 & 0 & 0\\
			0 & 1 & 0\\
			0 & 0 & 1
		\end{array}\right]x(k) + v(k).
	\end{align}
	The covariance of the Gaussian noise $v(k)$ is assumed to be $R=0.01I_3$. The other system parameters are set
	to $T = 1$s, $\alpha = 0.01$, and $\sigma_m^2
	= 10$. In this example,
	the original $(C,A)$ in \cite{singer1970estimating} is already in the form of Wonham decomposition. By Theorems \ref{th-decouple}
	and \ref{th-anaiff},
	the conditions $q_1>0, q_2>0$ and $q_3>0$ are sufficient to guarantee the MS detectability of the system.
	
	\begin{figure}[!]
		\centering
		\includegraphics[scale=0.62]{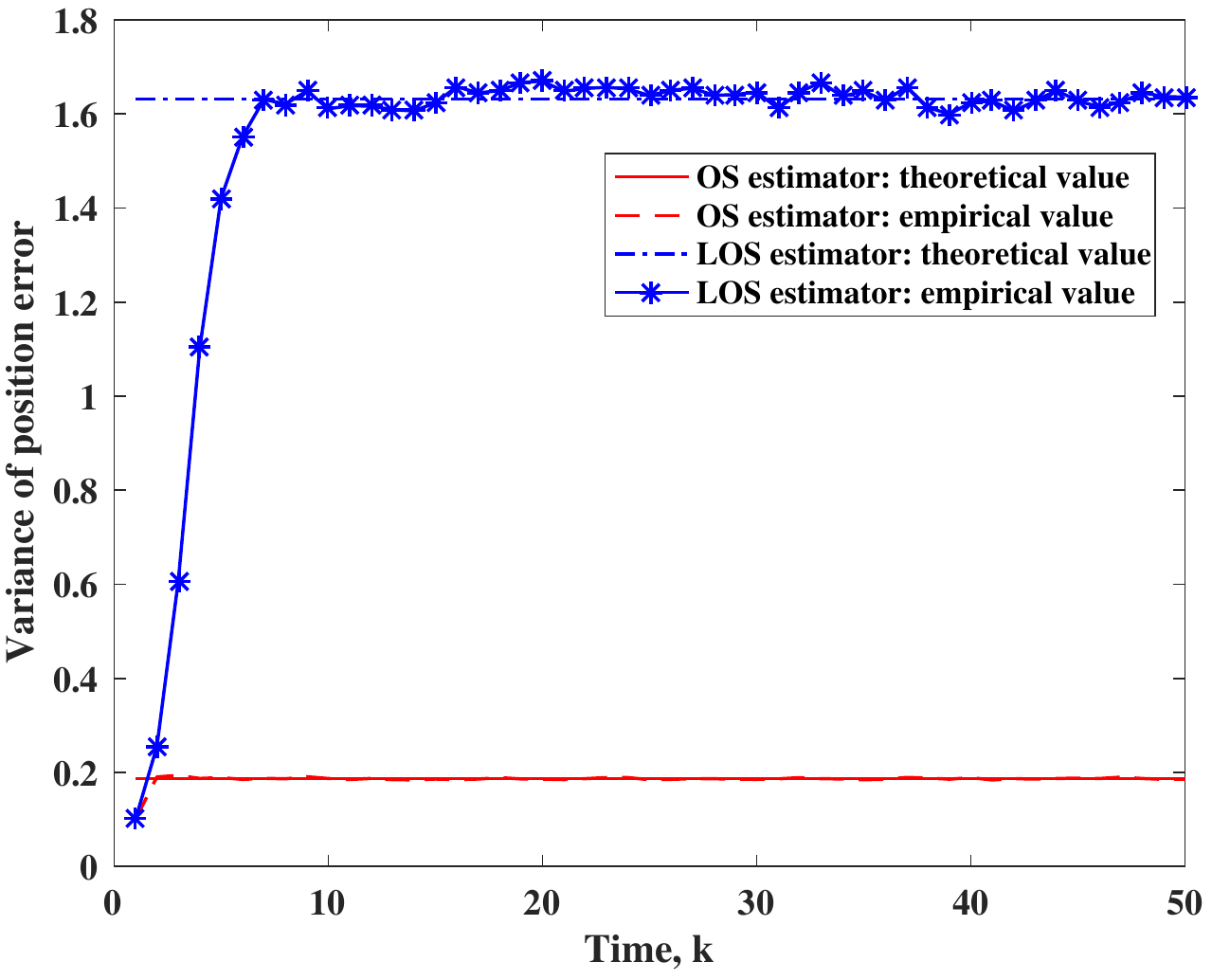}
		\caption{The target position error variance of the OS and LOS estimators. The theoretical value for OS estimator is the $(3,3)$th element in the matrix $\sum_{i=1}^{N}Y_i$, and the theoretical value for LOS one is the $(3,3)$th element in the matrix $\lim\limits_{k\to\infty}\sum_{i=1}^{N}\Sigma_i(k)$.}
		\label{fig_var}
	\end{figure}
	
	\begin{figure}[!]
		\centering
		\includegraphics[scale=0.62]{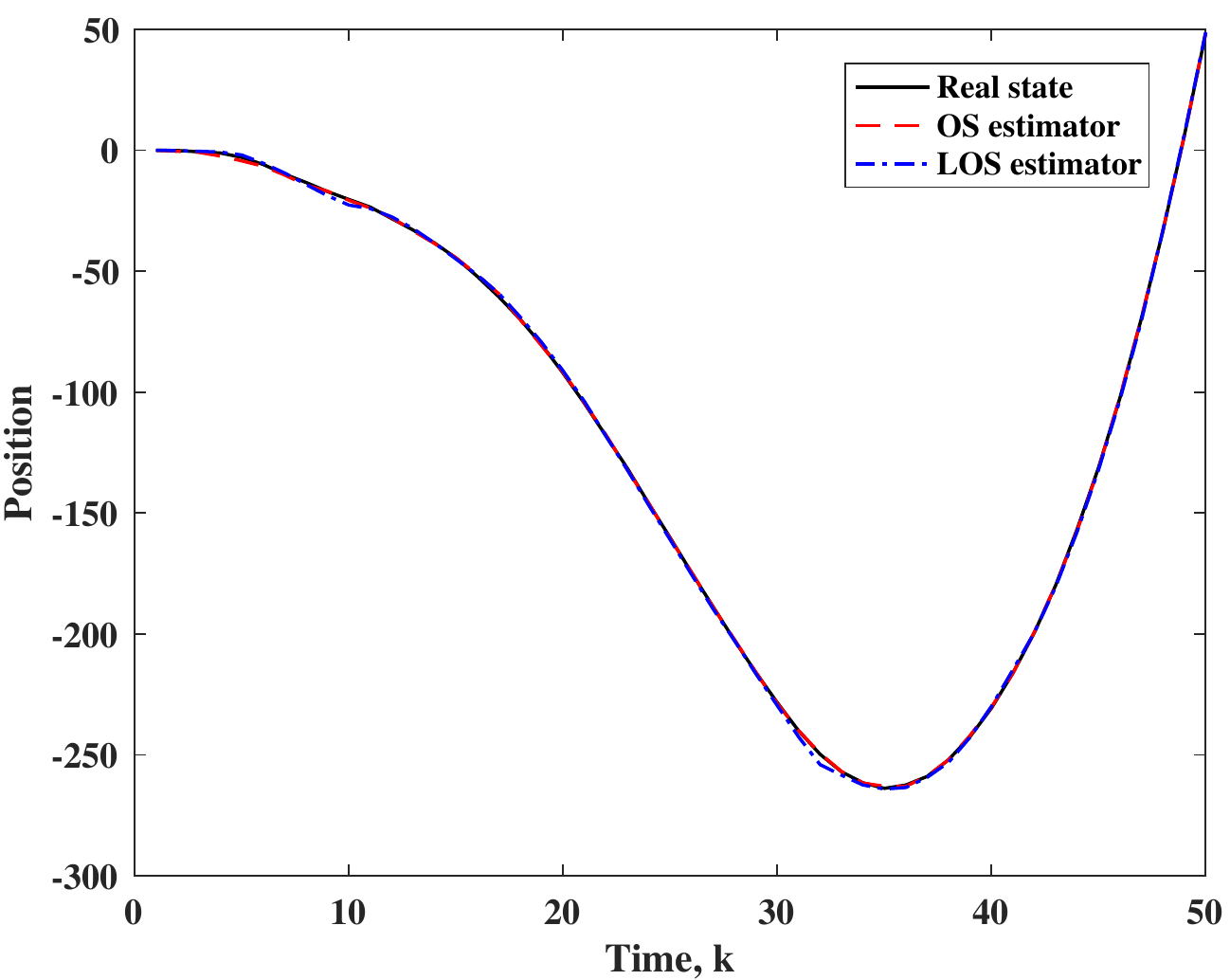}
		\caption{A realization of target position estimates of the OS and LOS estimators.}
		\label{fig_tracking}
	\end{figure}
	
	\begin{figure}[!]
		\centering
		\includegraphics[scale=0.62]{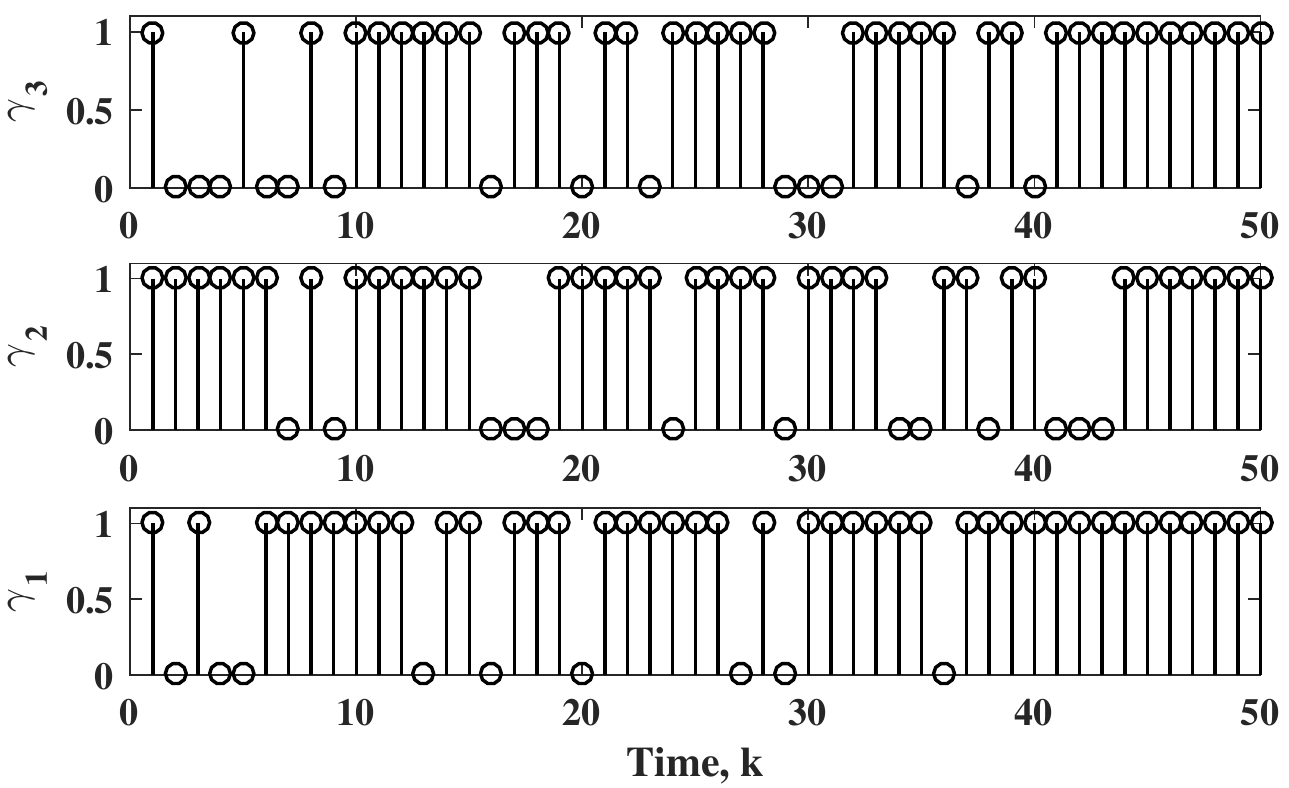}
		\caption{A sample path of packet drops. }
		\label{fig_gamma}
	\end{figure}
	
	Set the channel parameters as
	\begin{align*}
		p_1 = 0.20,\ \ \ p_2 = 0.30,\ \ \
		p_3 = 0.20, \\
		q_1=0.85,\ \ \ q_2=0.75,\ \ \
		q_3 = 0.80.
	\end{align*}
	For both OS and LOS estimators, we perform a Monte Carlo simulation with 50,000 trials over the time horizon of $k\in[1,\
	50]$ to show the estimation performance
	represented by the target position error
	variance. Fig. \ref{fig_var}
	shows the empirical variances of the OS and LOS estimators, of which both are close to their respective
	theoretical values. As expected, the performance of the LOS estimator is inferior to that of the globally optimal one that has much higher complexity. On the other hand, numerical results for the target position tracking shown in
	Fig. \ref{fig_tracking}, illustrate fairly good estimation performance for both estimators. Fig. \ref{fig_gamma} shows the corresponding packet drop sequences of sensor data for the tracking in Fig. \ref{fig_tracking}.
	
	\section{Conclusion}\label{sec-con}
	This paper studies the stationary state
	estimation over multiple Markovian
	packet drop channels. We have investigated
	the existence of the MS stabilizing solution,
	which is pivotal to the proposed
	estimator. It is shown that the known
	stabilizability condition in the existing literature is not necessary; only the controllability of the eigenvalues on the unit circle
	is required. In addition, a sufficient condition
	is derived showing that the MS detectability
	with multiple Markovian packet drop channels
	can be decoupled. Based on the decoupling method, a locally optimal stationary estimator with much lower complexity is proposed. In fact, the exponential complexity of the original optimal estimator is reduced to the linear complexity
	of the locally optimal estimator. Some analytic sufficient and necessary MS detectability conditions
	are also derived for the decoupled subsystems, each of which corresponds to the scenario of single Markovian packet drop channel. The results in this paper are potentially applicable to smart and
	optimal manufacturing in which a network of sensors collaboratively sense the interested process states \cite{qian2017fundamental}.
	
	\begin{ack}                               
		  The authors would like to thank the Associate Editor
		  and the anonymous reviewers for their constructive suggestions that
		  have improved the quality of this work.
	\end{ack}
	
	\appendix
	\section*{Appendix}
	The following technical Lemma will be useful in the proof of Theorem \ref{th-iff}.
	\begin{lemma}\label{le-matrix}
		The following two statements are true:
		\begin{itemize}
			\item[(i)] For any $A, B$ in $\mathbb{S}_+^n$, ${\rm tr}(AB) \ge 0$, and the equality holds
			if and only if $AB = 0$.
			\item[(ii)] For any $A\in\mathbb{C}^{m\times n}$, $AA^*=0$ if and only if $A=0$.
		\end{itemize}
	\end{lemma}
	
	\textbf{Proof of Theorem \ref{th-iff}}.
	\emph{Necessity:} It is obvious that the MS detectability for $(\mathbf H, \mathbf A,\mathbf p)$ is necessary. In
	order to prove that condition 2) is also
	necessary, assume on the contrary that
	condition 2) does not hold but the {CAREs} in
	(\ref{CARE}) has the MS stabilizing solution $Y$, implying that $\widetilde{\mathcal{L}}^*$ with $L_i = -K_i(Y)$ is a stable operator. Since condition 2) is false,
	from Definition \ref{def-uncontr} there holds
	\begin{align}\label{uncontr1}
		\mathcal{L}^*(V) = V,\ Q^{1/2}V_i = 0, \ \forall i\in\mathcal{N},
	\end{align}
	where
	$V=(V_1,\ldots,V_N) \in{\mathbb H}_+^n\backslash\{0\}$.
	In light of the optimal state
	estimator gains in (\ref{Kj})
	and the definition for
	$\mathcal{L}_j(\cdot)$ in
	(\ref{tpmlj}), {CAREs} (\ref{CARE})
	in Theorem \ref{th-estimator}
	can be rewritten as
	\begin{align}\label{care1}
		Y_j = \mathcal{L}_j(Y)
		-\sum_{i=1}^N p_{ij}\left[K_i(Y)H_iY_iA'
		-\mu_iQ\right].
	\end{align}
	Multiplying both sides of the
	above {CAREs} by $V_j$ from right,
	and applying the operations
	of trace and summation yield
	\begin{align}\label{tryv}
		\sum_{j=1}^{N}{\rm tr}(Y_jV_j) =&
		\sum_{j=1}^{N}{\rm tr}\{\mathcal{L}_j(Y)V_j\} + \sum_{j=1}^{N}
		\sum_{i=1}^{N}p_{ij}
		\mu_i{\rm tr}(QV_j) \notag\\
		& - \sum_{j=1}^{N}\sum_{i=1}^{N}
		p_{ij}{\rm tr}\left(K_i(Y)H_iY_iA'V_j\right).
	\end{align}
	By the inner product in
	(\ref{inner}) and the
	relationship in (\ref{adjoint}), we have
	\[
	\sum_{j=1}^{N}{\rm tr}\{\mathcal{L}_j(Y)V_j\} = \sum_{j=1}^{N}{\rm tr}\{Y_j\mathcal{L}^*_j(V)\}.
	\]
	Then combining (\ref{uncontr1}),
	we conclude that equality
	(\ref{tryv}) is equivalent to
	\begin{align*}
		\sum_{j=1}^{N}{\rm tr}(Y_jV_j) &=
		\sum_{j=1}^{N}\left[{\rm tr}(Y_jV_j)
		- \sum_{i=1}^{N}p_{ij}{\rm
			tr}\left\{K_i(Y)H_iY_iA'V_j\right\}\right]
	\end{align*}
	implying that $K_i(Y)H_iY_iA'V_j = 0\ \forall i,j\in\mathcal{N}$ by Lemma \ref{le-matrix} (i). Noticing the expression of $K_i(Y)$ in (\ref{Kj}), we further conclude $H_iY_iA'V_j=0$ by Lemma \ref{le-matrix} (ii), and thus
	$K_i(Y)'V_j=0$ $\forall i,j\in
	\mathcal{N}$. Therefore, by
	setting $L_i = -K_i(Y)$ in the
	operator $\widetilde{\mathcal{L}}^*$,
	we obtain
	\begin{align}
		\widetilde{\mathcal{L}}_i^*(V) &=
		\sum_{j=1}^{N}p_{ij}
		[A-K_i(Y)H_i]'V_j[A-K_i(Y)H_i]\notag\\
		& = \sum_{j=1}^{N}p_{ij}A'V_jA =
		\mathcal{L}_i^*(V) = V_i,
	\end{align}
	which means that $\widetilde{\mathcal{L}}^*(V)$ with $L_i = -K_i(Y)$ is unstable, contradicting the
	assumption on the MS stabilizing solution.
	This concludes the necessity of
	condition 2).
	
	\emph{Sufficiency:} 
	It suffices to show that
	$\rho(\widetilde{\mathcal{L}}^*)<1$
	under conditions 1) and 2).
	Assume on the contrary, $\rho(\widetilde{\mathcal{L}}^*)\geq 1$.
	Setting $L_i = -K_i(Y^+)$ implies that
	$\rho(\widetilde{\mathcal{L}}^*)=1$,
	by Lemma \ref{le-max} (i).
	Let $(\lambda=1, V)$ be an eigenvalue--eigenvector pair for  $\widetilde{\mathcal{L}}^*$ such that $\widetilde{\mathcal{L}}^*(V)=V$.
	Rewrite {CAREs} (\ref{CARE}) as
	\begin{align}
		Y_j^+ =& \sum_{i=1}^{N}p_{ij}
		\big\{[A-K_i(Y^+)H_i]
		Y_i^+[A-K_i(Y^+)H_i]'\notag\\
		& +
		\mu_i[K_i(Y^+)RK_i(Y^+)'+Q]
		\big\}.
	\end{align}
	The same manipulations as in
	(\ref{tryv}) and the adjoint relation $
	\sum_{j=1}^{N}{\rm tr}\{\widetilde{\mathcal{L}}_j(Y^+)V_j\}
	= \sum_{j=1}^{N}{\rm tr}\{Y_j^+\widetilde{\mathcal{L}}^*_j(V)\}$
	lead to
	\begin{align*}
		\sum_{j=1}^{N}{\rm tr}(Y_j^+V_j)
		=& \sum_{j=1}^{N}{\rm tr}\{Y_j^+\widetilde{\mathcal{L}}^*_j(V)\} \notag\\ +\sum_{j=1}^{N}
		\sum_{i=1}^{N}&p_{ij}\mu_i{\rm tr}\{K_i(Y^+)RK_i(Y^+)'V_j+QV_j\}.
	\end{align*}
	Since $\widetilde{\mathcal{L}}^*(V)=V$,
	the above equation implies that $QV_j=0$, and $K_i(Y^+)'V_j=0$, by again Lemma \ref{le-matrix}, further leading to ${\mathcal{L}^*}(V)=V$. This contradicts  condition 2)
	and concludes the sufficiency proof.
	The proof is now
	complete.	\hfill\rule{2mm}{2mm}
	
	
		\bibliographystyle{elsarticle-num}
		\bibliography{mybibfile}           
		
	
\end{document}